
%
\rightline {May, 1992}
\rightline {NSF-ITP-92-84, UPR-494-T, YCTP-P44-91}
\title {NONPERTURBATIVE STABILITY OF SUPERGRAVITY AND
 SUPERSTRING VACUA}
\author {Mirjam Cveti\v c\foot{email MIRJAM@PENNDRLS.UPENN.edu},
        Stephen Griffies\foot{email GRIFFIES@PENNDRLS.UPENN.edu} }
\address {Physics Department \break
     University of Pennsylvania, Philadelphia PA 19104-6396}
\andauthor {Soo-Jong Rey\foot{email
           SOO@SBITP.UCSB.edu, SOO@YALPH2.bitnet} }
\address { Institute for Theoretical Physics\break
University of California, Santa Barbara CA 93106\break
\& \break
Center for Theoretical Physics\break
Yale University, New Haven CT 06511 }

\abstract
{We investigate the possibility of false vacuum decay in $N=1$
supergravity theories. By establishing a
Bogomol'nyi bound for the energy density stored in the
domain wall of
the $O(4)$ invariant bubble, we show that supersymmetric vacua
remain absolutely stable against false vacuum decay
into another supersymmetric vacuum.
This conforms with and completes the previous
perturbative analysis of Weinberg.
Implications for
dynamical supersymmetry breaking and
decompactification instabilities in
superstring theory are discussed.
In addition, we show that there are no compact
static spherical domain walls.
}

\endpage

\chap{Introduction}
\REF\COL{S. Coleman, Phys. Rev. \bf D15 \rm (1977) 2929;
\bf D16\rm  (1977) 1248 (E);
C.G. Callan and S. Coleman, Phys. Rev. \bf D16 \rm (1977) 1762.}
The mechanism of false vacuum decay is presently
well understand\refmark{\COL}.
Quantum-mechanical tunnelling from the false vacua
to the true one proceeds via formation of a true vacuum
bubble inside the false vacuum
background. Energy gained by forming the true vacuum bubble
less the wall energy is transferred to the wall kinetic energy,
driving the wall expand asymptotically to the speed of light.

Coleman and DeLuccia\Ref\COLDL{S. Coleman and F. De Luccia,
Phys. Rev. \bf D21 \rm (1980) 3305.}
investigated effects of gravity to the
false vacuum decay. In particular, they found that a Minkowski
false vacuum cannot decay into an anti de Sitter (AdS)
true vacuum unless the matter vacuum energy difference
$\Delta V$ is sufficiently large:
$$
\Delta V \equiv V_{false} - V_{true} \ge {3 \over 4}\kappa
 \sigma^2 .
\eqn\cdlbound
$$
Here, $\kappa\equiv 8\pi G_N$ and
 $\sigma $ denotes the bubble wall energy per unit area.

The bound implies that tunnelling takes place only
if difference of the matter potential energy between
false and true vacua is large enough.
When  the  bound \cdlbound\
is saturated, the critical size of the bubble grows
indefinitely. However, since
the bubble energy density $\sigma$ remains finite,
it takes an infinite amount of energy
to create the true vacuum.  Therefore, in this limit
the tunnelling process shuts down.

The Coleman-DeLuccia bound can be generalized to false
vacuum decay between two AdS vacua with
corresponding matter vacuum energies
$ V_{true}<V_{false}<0$. In this case, by applying the
formalism of Ref.~\refmark\COL\ , we find that
$$ (\sqrt{-V_{true}}-\sqrt{-V_{false}})^2\ge
{3 \over 4}\kappa \sigma^2
\eqn\cdlboundII$$
must be satisfied in order for  a true vacuum
bubble to form.

Superstring theories might provide the first unified
theory of all known interactions \sl including quantized
gravity\rm. In realistic Calabi-Yau
compactifications, the four-dimensional
low-energy massless string modes are described by
an $N=1$ supergravity. In such models, it is of foremost
importance to find and understand a mechanism to break
local supersymmetry at a low-energy scale. The folklore
theorem, which is partially proven in an extended spacetime
supersymmetry\Ref\Dixonbanks{L. Dixon and
T. Banks, Nucl. Phys. \bf B307 \rm (1988) 93.}, tells us
that spacetime supersymmetry cannot be continuously broken
within a class of classical vacua. We must then rely only on
nonperturbative stringy effects to break the supersymmetry.
Several sources of dynamical supersymmetry breaking have been
proposed; for example, hidden sector gluino
condensation\Ref\GLUINO{J.P. Derendinger, L.E. Ib\'a\~nez and
H.P. Nilles, Phys. Lett. \bf 155B \rm (1985) 65;
M. Dine, R. Rohm, N. Seiberg and E. Witten, Phys. Lett. \bf
156B \rm (1985) 55.}, gravitational instantons\Ref\GRAVINST{
K. Konishi, N. Magnoli and H. Panagopoulos, Nucl. Phys. \bf
B323 \rm (1989) 441.}\Ref\NEW{S.-J. Rey,
\sl Abelian Heterotic Instanton in Superstring
Theory, \rm YCTP-P18-92 preprint (1992).}
and axionic instantons\Ref\REY{
S.-J. Rey, Phys. Rev. \bf D43 \rm (1991) 526:
\sl String Theory and Axionic Strings and Instantons, \rm talk
given at Particle and Fields `91, Vancouver, Canada (1991).}.
However, a generic difficulty associated with such
nonperturbative mechanisms is the existence of multiple local
vacua, but only some of them are supersymmetric.
Eventually, we would like to explain issues such as
phenomenologically viable gauge groups, matter contents and
dynamics of supersymmetry breaking within this
framework. On the other hand, since supersymmetric vacua
seems to proliferate, we would also like to understand
whether
a chosen, phenomenologically promising superstring vacuum
remains semiclassically stable. If one happens
to sit in a local supersymmetric
vacuum, can one tunnel to another minimum which also
preserves supersymmetry?

In this paper, we point out that such quantum tunnellings
are absolutely impossible. More specifically, vacuum decay
from a supersymmetric Minkowski vacuum to an AdS
supersymmetric vacuum is not possible at all.
In particular, this applies to the case of tunnelling in the
gauge neutral moduli directions when the non-perturbative
potential is turned on.\foot{This does not, however,
preclude the possibility of \sl thermal \rm transition over the
compactification potential barrier. Indeed, this is the most
relevant dynamics in the context of stringy
cosmology\Ref\NEW{T.W.B. Kibble and S.-J. Rey,
work in progress (1992).}.}
On the other hand, we prove this result only in the matter
sector of chiral superfields, and do not
address the issue of non-zero background gauge fields either,
even though we suspect a similar conclusion can be drawn
in these more general cases.

Our technical assumption is that the tunnelling  among
string vacua is well described by four-dimensional $N=1$
supergravity theory with a nonzero superpotential for the
matter sector responsible for the tunnelling. In particular,
this sector
consists of moduli fields with the  nontrivial superpotential
induced by the aforementioned nonperturbative effects.
However, our discussion of vacuum tunnelling
is quite generic. Thus, our final conclusions shoule apply
to tunnelling in not only superstring theory but also
in any $N=1$ supergravity theory.  In the discussion of
superstring decompactification instability, we will also
learn that one important stringy ingredient,
the spacetime modular invariance, plays a crucial role.

Stability of supersymmetric vacua in the  context of
supergravity grand unified
theories was investigated earlier by Weinberg\Ref\WEIN{S.
Weinberg, \sl Phys. Rev. Lett.\bf 48 \rm (1982) 1776.}
with a conclusion that supersymmetric vacua are stable
against a false vacuum decay.
However, his analysis was done
only in the leading (perturbative) order in a
$G_N$ expansion.
The result presented in this
paper establishes the same
conclusion, however, to all orders in $G_N$.

The paper is organized as follows: in Chapter 2
we prove by using the  $O(4)$ invariant Ansatz
that the energy density
of the domain wall bubble is bounded from below.
The only bubble that can form  corresponds to an
infinite radius bubble and we present its
explicit form.
 In Chapter 3, we discuss the tunnelling among the
superstring vacua; conclusions are given in
Chapter 4. Technical details within the context
of general
spherically symmetric domain wall configurations
are discussed in the Appendix.

\chap{Minimal Energy Configurations of the
  Vacuum Bubble}

\section{Lagrangian}

We consider four-dimensional $N=1$ local supersymmetry with
one chiral superfield $T$. The analysis can be
straightforwardly generalized to multi-matter superfields.
The bosonic sector of the  four-dimensional
$N=1$ supergravity Lagrangian reads
$$
e^{-1} L = -{1 \over 2 \kappa} R + K_{T \bar T} g^{\mu \nu}
\nabla_\mu T \nabla_\nu \bar T - V(T, \bar T)
\eqn\lagran
$$
in which the supergravity scalar potential $V(T, \bar T)$
is defined as
$$
V \equiv e^{\kappa K} [ K^{T \bar T} |D_T W|^2
- 3\kappa|W|^2]
\eqn\potential
$$
here
$e = |detg_{\mu \nu}|^{1 \over 2},
K(T, \bar T) =$ K\"ahler potential and $D_{T}W \equiv
e^{-\kappa K} (\partial_T e^{\kappa K}W)$.
\foot{We do not choose the commonly used  K\"ahler gauge
which introduces the potential function\Ref\WB{ Wess and
Bagger, \sl Supersymmetry and Supergravity \rm ,
2nd edition, Princeton 1992.}
$G(T,\bar T) = K(T,\bar T) + ln|W(T)|^{2}$ since it
is not adequate for situations in which the superpotential
is allowed to vanish.}
Newton's constant appears consistently in the combination
$\kappa = 8 \pi G_{N}$.

Supersymmetry preserving minima of the scalar potential
\potential\ satisfy
$D_{T}W=0$. This in turn implies (see eq.\potential\ )
that the supersymmetry preserving vacua have
either zero vacuum energy (Minkowski space-time)
when $ W=0$, or constant negative vacuum energy
$- 3\kappa e^{\kappa K} |W|^2$  (AdS space-time) when
$W\not=0$.  Thus, the tunnelling process between
supersymmetric vacua corresponds to tunnelling either
between Minkowski and AdS space-times or
between two AdS space-times of different cosmological
constants.

In superstring theories,
the scalar field $T$ corresponds to a modulus field
arising from
compactification, and its non-perturbatively induced
superpotential $W$ is assumed to reflect
the underlying target space modular invariance\Ref\CFILQ{M.
Cveti\v c, A. Font, L. E. Iba\~ nez, D. L\" ust and
F. Quevedo, Nucl. Phys. \bf B361 \rm (1991) 194.}
 under the $PSL(2,{\bf Z})$
duality transformations:
$$T\rightarrow{{aT-ib}\over{icT+d}}  \ \ \ ,
 ad-bc=1\ \ ,
 \{a, b, c, d\}\in {\bf Z}.\eqn\modtrans
 $$
In this case\refmark\CFILQ ,
$
K = - 3 \kappa\ln [(T + \bar T)]$.
The superpotential $W$ is a modular  function  of weight
$-3$ under  $PSL(2, \bf Z)$ defined in the fundamental
domain $\cal D$ of the $T$-field.
The most general form of the superpotential is
           $P(j(T))\cdot \eta^{-6}(T)$
in which $\eta (T)$ is the Dedekind function: a modular
function of weight $1/2$.  $P(j(T))$ is a rational
polynomial of $j(T)$: a modular-invariant function
\Ref\SCH{B. Schoeneberg, \sl Elliptic Modular Functions,
\rm Springer-Verlag 1974.}.
One of the simplest choices for a
modular invariant superpotential is:
$$W(T) = j(T)\, \eta^{-6}(T) .\eqn\sdual$$
In this case the  scalar
potential for the  $T$ field has two, supersymmetric
minima at $T=1$ ($V<0$)
and $T=\rho \equiv e^{i\pi/6}$ ($V=0$), both of which
lie on the boundary of the fundamental domain ${\cal D}$.
This is an example of two supersymmetric non-degenerate
superstring vacua.

\section{Bogomol'nyi
Bound For  The Bubble Domain Wall Energy}

In this chapter, we
show that for $N=1$ supergravity,
the bubble wall energy density for a given time slice
is finite and bounded below by
a `topological charge'.  We show that
the  bubble wall of minimum energy density
separates two supersymmetric vacua.  However, analysis
of the expressions for the first order equations arising
from the saturation of this  action bound will reveal
that the bound is
saturated \sl only for a wall of an infinite radius\rm.
This result implies that  there is \sl no \rm vacuum
tunnelling between supersymmetric vacua in the theory.
Consequently, this
result implies that there are no compact
(finite radius), static, supersymmetric,
spherical  domain wall solutions in $N=1$ supergravity.
Thus, all supersymmetric vacua are energetically
 degenerate.
Note that in the discussion, what we mean by `false'
and `true' vacuum is in reference to their relative
matter energy density as determined by the scalar
potential.  The essence of our result is that
there are no `false' supersymmetric vacua in
$N=1$ supergravity.  The reason is that gravity and
matter exactly balance their energies such that all
supersymmetric vacua become degenerate.

We study the bubble formation
by using   an $O(4)$ symmetric Ansatz\refmark{\COL}
for a bounce solution interpolating between the
true(AdS) and false(Minkowski or AdS) vacuum.  The
metric for this Ansatz in Euclidean space
is\refmark{\COLDL  }
$$ \eqalign{
ds^{2}& = d\xi^{2} + R(\xi)d\Omega_{3}^{2}\cr
&=B(\xi')(d\tau^2+dr^2+r^2d\Omega_2^2)}
\eqn\ofourm$$
where $\xi$ is the Euclidean radial distance from
an arbitrary origin and $\xi '^2=\tau^2+r^2$.
The second line of \ofourm\ follows after a
redefinition of the radial coordinate $\xi$ into
$\xi'$: $d\xi '/\xi '=d\xi/\sqrt{R(\xi)}$.
Note that only with  coordinate
  $\xi '$   we can clearly attach the meaning to
$r=\sqrt{x^2+y^2+z^2}$ as the radius of the two sphere.
 The classical evolution of the materialized bubble
is described by the Wick rotation back to Minkowski
space-time,
\ie , by changing the Euclidean time $\tau$ back to
Minkowski time $t$.

It is most convenient to study the energy density of
the bubble wall at the moment of its actual formation, \ie
at Euclidean time  $\tau =0$. At this moment the bubble is
instantaneously at rest\refmark{\COL \COLDL};
the time derivative of the matter field
$\partial_\tau T\equiv
(\tau /\xi ')
\partial_{\xi'}T
$  and the metric coefficient
$\partial_\tau B\equiv
(\tau /\xi ')
 \partial_{\xi '}B\
$
both vanish at $\tau=t=0$.
It turns out (see section 2.3)
that for the minimal energy configuration of the bubble,
the metric coefficient $B$ and the matter field $T$ satisfy
first order differential equations. This in turn justifies
the choice that at $\tau =t=0$ the matter field $T$
and the metric coefficient $B$
of the minimal energy configuration
are only  functions of $r=\sqrt{x^2+y^2+z^2}$, \ie , the
radius of the  two sphere.
At the moment of the actual bubble formation
one is thus working with a specific
spherically symmetric metric
Ansatz:\foot{We calculate explicitly in the
Lorentzian instead of Euclidean signature. The
conclusion about the positive minimal energy stored
in the bubble wall or the minimal action theorem for
the bounce solution are equivalent
since time independence is assumed throughout. }
 $$
 ds^2 = B(r)(dt^2 -  dr^2 - r^2 d \Omega_2^2)
 \eqn\metric
 $$
In the Appendix a  study of a  general spherically symmetric
 $O(3)$  configuration with  the metric Ansatz:
 $ ds^2 = B(r) dt^2 - A(r) dr^2 - C(r) d \Omega_2^2$
is presented. There, for the minimal energy configuration,
one is able to see that with $A(r)=B(r)$,
 $C(r)=B(r)r^2$, thus
reducing the general spherically symmetric
configuration  to the metric \metric\ of
the bounce at the time of actual bubble creation.

For the purpose of studying  the minimal energy
configuration of the bubble wall energy density
we introduce supersymmetry charge density:
\foot{Our conventions are the following:
 $\gamma^{\mu}=e^{\mu}_{a}\gamma^{a}$ where
$\gamma^{a}$ are the usual Dirac matrices satisfying
$\{\gamma^{a},\gamma^{b}\}=2\eta^{ab}$;
$e^{a}_{\mu}e^{\mu}_{b} = \delta^{a}_{b}$; $a=0,...3$;
$\mu=t,x,y,z$; $\overline{\psi} = \psi^{\dagger}\gamma^{t}$;
$(+,-,-,-)$ space-time signature.}
$$
Q[\epsilon'] = 2 \int_{\partial \Sigma} d \Sigma_{\mu \nu}
(\bar \epsilon' \gamma^{\mu \nu \lambda} \psi_\lambda)
\eqn\susytrans
$$
where $\Sigma$ is a space-like hypersurface enclosing the
bubble wall. Here, $\epsilon'$ is a commuting Majorana
spinor and $\psi_{\rho}$ is the spin $3/2$ gravitino field.
Taking a supersymmetry variation of $Q[\epsilon']$ with
respect to another commuting Majorana spinor
$\epsilon'$ yields
$$
\delta_{\epsilon} Q[\epsilon'] \equiv \{Q[\epsilon'],
\bar{Q}[\epsilon]\}
= \int_{\partial \Sigma}N^{\mu \nu} d\Sigma_{\mu \nu}
= 2\int_{\Sigma}\nabla_{\nu}N^{\mu \nu} d\Sigma_{\mu}
\eqn\localchargevariation$$
where we introduced the generalized Nester's form\Ref
\NESTER{J. M. Nester, Phys. Lett. \bf 83A \rm
(1981) 241.}
$$N^{\mu \nu} = \bar \epsilon'\gamma^{\mu \nu \rho}
\hat\nabla_{\rho} \epsilon\, \ . \eqn\nesterform
 $$
The supercovariant derivative appearing in Nester's form is
$$\hat\nabla_{\rho}\epsilon \equiv
\delta_{\epsilon}\psi_{\rho} =
[2\nabla_{\rho} + ie^{\kappa K / 2}(WP_{R} +
\bar{W}P_{L})\gamma_{\rho}
 - Im(K_{T}\partial_{\rho}T)\gamma^{5}]\epsilon\
\eqn\gravitinotrans$$ where the gravitational
derivative acting on a spinor is
$ \nabla_{\mu}\epsilon = (\partial_{\mu}
  + {1\over2}\omega^{ab}_{\mu}\sigma_{ab})\epsilon$.
In \localchargevariation\
the last equality follows
from Stoke's law.

We can describe the energy  stored in the bubble wall
(or equivalently the minimal action stored in the wall
of the bounce solution) using a thin wall
approximation\refmark{\COLDL} .
Such an approximation is valid in the case when the
radius $R$ of the bubble is much larger than its thickness
$2 \Delta R$, and becomes exact when $R\to \infty$.
The boundary condition on the metric coefficient is
$B(r=R)=1$, which serves as a suitable choice for
normalizing the metric. This in turn defines
the surface of the large radius bubble to be
$ 4\pi R^2$.
In the thin wall approximation, in the region
with $r\sim R$,  the metric coefficients do not change
appreciably over the range of the domain wall.
The boundary $\partial \Sigma$ are two boundaries of
two-sphere, one  at $ R -\Delta R$ and the other
at $ R+ \Delta R$, with the constraint that
$R\gg 2 \Delta R$.
In this limit, the spherical domain wall approaches
the planar domain wall\Ref\CGR{M. Cveti\v c, S. Griffies,
S.  J. Rey, \sl Static Domain Walls in $N=1$ Supergravity
\rm, UPR-474-T, YCTP-P43-91 (January 1992),
Nucl. Phys. \bf B \rm in press.}.

The volume integral in eq.\localchargevariation\
yields\refmark{\CGR}
$$
2\int_{\Sigma}\nabla_{\nu}N^{\mu \nu} d\Sigma_{\mu}=
\int
[-\delta_{\epsilon'}\psi^\dagger_i g^{ij}
\delta_{\epsilon}\psi_j +
K_{T \bar T}\delta_{\epsilon'}\chi^\dagger
\delta_{\epsilon}\chi]
{B(r)}r^2drd\Omega_{2}.
\eqn\volumeintegral$$
where $\delta_{\epsilon}\psi_{i}$ and
$\delta_{\epsilon}\chi$ are the supersymmetry variations
of the fermionic fields in the bosonic backgrounds.
Upon setting $\epsilon' = \epsilon$ the expression
\volumeintegral\
is a positive definite quantity which in turn (through
 eq.\localchargevariation\ )
yields a bound $\delta_{\epsilon} Q[\epsilon] \ge 0.$

Analysis of the surface
integral in \localchargevariation\
yields two terms: $(1)$ The ADM mass of the configuration,
denoted $ 4\pi  R^2 \cdot \sigma$
and $(2)$ The topological charge,
denoted $4\pi  R^2 \cdot \cal{C}$.
Here, $\sigma$ and $\cal{C}$ denote the ADM mass density
and the topological charge density
of the bubble wall. As explicitly shown in the Appendix,
the minimum topological charge density, which corresponds
to a supersymmetric configuration,  is given by
$$
\eqalign{
|\cal{C}| = & \, 2 \, |\, (  \zeta
|We^{\kappa K \over 2}|)_{r=R+\Delta R}
-( \zeta
|We^{\kappa K \over 2}|)_{r=R-\Delta R}\,|
\cr
=&{2\over {\sqrt{3\kappa}}}|
 \zeta_{R-\Delta R}
\sqrt{-V_{true}}
- \zeta_{R+\Delta R}
\sqrt{-V_{false}}|}
\eqn\topologicalcharge$$
where $\zeta=\pm 1$.  $\zeta_{R-\Delta R}=-
\zeta_{R+\Delta R}$ if $W$ goes through 0 somewhere as $r$
traverses an interval $(R-\Delta R, R+\Delta R)$ and
   $\zeta_{R-\Delta R}=\zeta_{R+\Delta R}$  otherwise
(see discussion of the minimal energy solution in the
following.)
The second line in eq.\topologicalcharge\ follows from the
properties of the supersymmetric vacua, namely, as we
have shown in section 2.1 for the supersymmetric
minimum, $V=-3\kappa |W|^2e^{\kappa K}$.

Positivity of the volume integral translates into the bound
$$
\sigma \ge |\cal{C}|
\eqn\localbound$$
which is saturated if and only if $\delta_\epsilon
Q[\epsilon]=0$.
Saturation of this bound implies the bosonic
background is supersymmetric and
$\delta\psi_{\mu} = 0$ and
$\delta\chi = 0$.
(see eq.\volumeintegral\ ).
In particular for the tunnelling from  Minkowski
($V_{false}=0$ since
$W_{R+\Delta R}=0 $)
to AdS
($V_{true}\neq 0$ since
$W_{R-\Delta R}\neq 0 $)  space-time, the
inequality \localbound\ implies:
$${3\over 4}\kappa
\sigma^2\geq
3\kappa
|W e^{\kappa K}|^2_{R-\Delta R}
\,\, = \,\, - V_{true}.
\eqn\minkowskibound $$
On the other hand,
the tunnelling from  AdS
($V_{false}\neq 0$ since
$W_{R+\Delta R}\neq 0 $)
to AdS
($V_{true}\neq 0$ since
$W_{R-\Delta R}\neq 0 $)  space-time  when $W$ does
not go through zero in between, the
inequality \localbound\ implies:
$${3\over 4}\kappa
\sigma^2\geq
3\kappa
(|W e^{\kappa K}|_{R+\Delta R}
-|W e^{\kappa K}|_{R-\Delta R})
=(\sqrt{-V_{true}}
-\sqrt{-V_{false}})^2.\eqn\adsbound$$

The two inequalities, \minkowskibound\ and \adsbound\ ,
are the central results of this paper.

Notice that the positive energy bounds
(\minkowskibound\ and \adsbound ) for the minimal
energy density  of the thin domain wall
have precisely the opposite inequality sign  as
eqs.\cdlbound\  and  \cdlboundII of the Coleman-DeLuccia
bound for the existence of a bubble instanton!
In other words, vacuum tunnelling is not allowed in
$N=1$ supergravity since the available vacuum energy
difference is not sufficient to materialize the
tunnelling bubble. For the marginal case\foot{
In the case that the bubble forming between the two
AdS space-times has $W$ going through 0 somewhere
in between, the supersymmetric configuration satisfies
${3/ 4}\kappa
\sigma^2=
3\kappa
(|W e^{\kappa K}|_{R+\Delta R}
+|W e^{\kappa K}|_{R-\Delta R})
=(\sqrt{-V_{true}}
+\sqrt{-V_{false}})^2$
and so the Coleman-DeLuccia bound is \sl never \rm
saturated; i.e. tunnelling is super-suppressed.
Therefore, we do not discuss this case further. }
when the bounds \minkowskibound\ and \adsbound\
are saturated, such a domain wall configuration
is supersymmetric.
However, even in this case, since the matter energy
gain by tunnelling to the true supersymmetric vacuum
is precisely cancelled by the bubble wall energy plus
the gravitational energy of the AdS space-time inside
the bubble wall, the result is an infinitely large
radius of the critical tunnelling bubble as we will
show in the next subsection.
Thus, the tunnelling is completely suppressed even in
this marginal case.

\section{Self-dual Equations}

In this section we analyze the first order self-dual
equations which the matter, geometry and spinor must
satisfy in order that the minimum energy density in the
configuration of the bubble to be realized.
We will find that these equations can be
satisfied only for a sphere of an \sl infinite \rm
radius.

We take the coordinate system specified by \metric\ .
The metric component $B(r)$
and matter field $T(r)$ depend only on the radius $r$
of the two sphere. However, we allow for
 the full spatial dependence of
the Majorana 4-spinor parameter $\epsilon$ .
Again, there is no time dependence.

This calculation involves an analysis of
the first order Bogomol'nyi equations
$\delta_{\epsilon} \chi = 0$ and $\delta_{\epsilon}
 \psi_\mu = 0$ which are necessary conditions for a
supersymmetric bosonic configuration.
Technical details for the derivation
of these equations in the spherical frame
are presented in the Appendix. Setting $\delta_{\epsilon}
\chi = 0$  yields the equation:
$$
K^{T\bar T}e^{ {\kappa K \over 2} } (D_{T}{\overline
W}P_R + D_{T} W P_L)\epsilon (r,\theta,\phi)
+ { i \over \sqrt B}\gamma^3
 ( \partial_r T \, P_L +
\partial_r \bar T \, P_R) \epsilon (r,\theta,\phi)
 = 0 \eqn\modulino
$$
where $P_{R,L}=(1\pm i\gamma_5)/2$.
This equation is satisfied if and only if:
$$
\partial_{r}T(r) = i e^{i \Delta (r)} \sqrt B
K^{T \bar T}
e^{\kappa K/2} D_{\bar T} \bar W. \eqn\tofr
$$
The two complex components of the Majorana spinor
$\epsilon^{T} = (\epsilon_1,\epsilon_2,\epsilon_2^*,
-\epsilon_1^*)$
must also be related through
$$(\epsilon_1, \epsilon_2) =  (\epsilon_1^o,
e^{i \Delta (r)} \epsilon_1^{o*})\eqn\epsilont$$
where
$\epsilon_{1}^{o} = \epsilon_{1}^{o}(r,\theta,\phi)$.

Similarly, $\delta_{\epsilon}\psi_{\mu} = 0$ gives
the following four equations (See Appendix for the more
general spherical metric Ansatz.):
$$\eqalign{
\gamma^o \, [\gamma^3 \, { {1\over 2} \partial_r{\ln B}}
+ i\kappa {\sqrt B} e^{\kappa K/2} (WP_R+\bar WP_L)
 ]\epsilon (r, \theta , \phi ) = 0, \cr
\{2 \partial_r - i \gamma^3 \kappa
{\sqrt B} e^{\kappa K/2} (WP_R+\bar WP_L)
 - \gamma^5\kappa Im (K_T \, \partial_rT )\}
\epsilon (r, \theta , \phi ) = 0,  \cr
\{ 2 \partial_\theta - \gamma^1 [\gamma^3
{\partial_r {\sqrt{B r^2}}
\over \sqrt B} + i \kappa {\sqrt{Br^2} } e^{\kappa K/2}
(WP_R+\bar WP_L)]
   \} \epsilon (r, \theta , \phi )
= 0,\cr
 \{ 2 \partial_\phi
- \gamma^2 [ \gamma^3 \sin \theta  ({ \partial_r
{\sqrt{Br^2}} \over
\sqrt B}  + i\kappa  \sqrt{B r^2} e^{\kappa K/2}
(WP_R+\bar WP_L)) + \gamma ^1 \cos \theta ]\}
\epsilon (r, \theta , \phi ) = 0.}
\eqn\gravitino
$$

We now show  that the vacuum bubble is supersymmetric only in
an infinite radius limit. The $t$-component of eq.\gravitino\
  gives  an equation for the metric component $B$
$$
 \partial_r({1\over{\sqrt B}})
= -i\kappa
e^{i\Delta (r)} e^{\kappa K/2}{\overline W}.
\eqn\tcomp
$$
The $r$-component
eq. \gravitino\  constrains the
radial dependence of the spinor:
$$
(\epsilon_{1}, \epsilon_{2}) =
  B^{1/4}  e^{i \Delta (r)/2}
\left(\tilde\epsilon_1(\theta , \phi ),  \tilde\epsilon_2
(\theta ,\phi ) \right), \eqn\eps$$
with
$$\tilde\epsilon_1(\theta ,\phi)
 = \left(\tilde\epsilon_2 (\theta ,\phi)\right)^{*}.
\eqn\rcomp$$
In addition, the phase $\Delta (r)$ satisfies
$$\partial_{r}\Delta (r) =
- \kappa Im(K_T\partial_rT). \eqn\deltar$$

With the expression for the metric
\tcomp\ , equation for the
$\theta$-component of \gravitino\ can be written as
$$
[2 \partial_\theta - \gamma^1 \gamma^3] \epsilon = 0
\eqn\thetacomp
$$
while the $\phi$-component is
$$
[2 \partial_\phi+
\gamma^1 \gamma^2 (-2 \sin \theta \partial_\theta
+ \cos \theta)] \epsilon = 0.
\eqn\phicomp
$$
Now eq. \phicomp\
can only be solved with a spinor of the form
$$
(\tilde\epsilon_1(\theta,\phi),\tilde\epsilon_2
(\theta ,\phi ))=e^{i\sigma_2\theta/2}
e^{i\sigma_3\phi/2}(a_1, a_2)
\eqn\epsII$$
with $a_1$ and $a_2$ constant complex coefficients
and $\sigma_{2,3}$ the usual Pauli matrices.

However, the spinor form \epsII\  is
incompatible with the constraint \rcomp .
Namely,  for any value of the angles $\theta$ and $\phi$
there are  {\it no} constant coefficients $a_1$ and $a_2$
in \epsII\ for which the constraint \rcomp\ can be satisfied.
On the other hand, this argument can be evaded in an infinite
radius limit. Indeed, there exists a planar
static supersymmetric
\REF\CG{M. Cveti\v c, S. Griffies, \sl Gravitational Effects
in  Supersymmetric Domain Wall Backgrounds \rm , UPR-503-T
(April 1992), Phys. Lett \bf B \rm in press.}
domain wall\refmark{\CGR ,\CG}, which may be identified
with the vacuum bubble in the
limit as the bubble radius of curvature $R\to\infty$.

We now show that the two constraints \epsII\ and
\rcomp\ become compatible in the limit as
the radius of curvature of the bubble $R\to \infty$.
We interpret this result as an infinite radius
static spherical domain  wall configuration which
saturates the Bogomol'nyi bound (eq.\localbound\ ).
For a nearly flat bubble wall, it is sufficient to fix a
value of the polar angles $\theta \approx \theta_0$
and  $\phi \approx \phi_0$.
Thus, one is studying a local region of the bubble wall
corresponding to an infinitesimal solid angle
$\Delta \Omega_2=
\sin\theta_0(\Delta\theta\Delta\phi $).\foot{When
$\theta_0=0$ or
$\theta_0=
\pi$ the two constraints can be simultaneously
satisfied for any value of $\phi$,
however, the solid angle is still infinitesimal, \ie ,
$\Delta\Omega_2=2\pi(\Delta\theta)^2$.}
In the leading order in $R \to \infty$,
\epsII\  and  \rcomp\ are compatible.  The
equations  of motion for the
matter field $T(r)$
(eq.\tofr) and the metric coefficient $B(r)$ (eq.\tcomp)
can be written as:
$$\eqalign{\partial_r T(r) &= \zeta \sqrt{B}|W|
e^{\kappa K/2}K^{T \bar T}
{D_{\overline{T}}\overline{W}\over \overline{W}},\cr
  \partial_r({1\over {\sqrt B}})&=
\kappa\zeta |W|e^{\kappa K/ 2}.}
\eqn\summary$$
The superpotential is related to the phase $\Delta(r)$
through $W=-i\zeta e^{i\Delta}|W|$  with $\zeta=\pm 1$
as follows from  eq.\tcomp\ ; \ie , the right hand side
 of eq.\tcomp\ should be real in order for the metric
to be real. Note that $\zeta$ can change
sign only at a point where $W$ vanishes.

The constraint on the phase of $W$ along with  the
first of \summary\
imposes the geodesic equation\refmark\CGR :
$$
Im(\partial_{r}T{D_{T}W\over W}) = 0.
\eqn\geodesic$$
This result implies that in the limit $\kappa\to 0$,
 $W(r)$ lies in the $W$ plane
 on a straight line that extends
through the origin.

We repeat that the matter field and the metric components
satisfy eqs.\summary\ only in the limit
as the bubble wall becomes flat: $R\to\infty$.
{}From \summary\ we can now deduce the qualitative features
of the  metric.

\section{Explicit Solutions in $R\to\infty$ Limit}

In general, the matter field $T(r)$  which
satisfies  the first equation in \summary\
specifies the thickness of the domain wall. It
follows the usual kink path, approaching exponentially
fast the supersymmetric `false' vacuum at
$r \sim R+\Delta R$ and the supersymmetric `true'
vacuum at $r\sim R-\Delta R$.

On the other hand,
the metric coefficients in general change relatively
slowly (see the form of the explicit solutions below).
In order to find the explicit form of the solution for
the metric, we impose  the following boundary condition
at the location of the domain wall: $B(R)=1$.
This condition turns out to be a suitable choice for
defining the metric in the presence of the bubble.

At this point we relate
the metric \metric\ with the metric of the
planar domain wall.
For the planar domain walls (in $(x,y)$ plane),
 the  line  element is specified as\refmark{\CGR ,\CG}:
$$ d s^2=B(z)(dt^2-dr^2-r^2d\Omega_2^2)
\eqn\pllineel$$
where $B(z)$ and $T(z)$
satisfy\refmark{\CGR ,\CG}:
$$\eqalign{\partial_z T(z) &= \zeta \sqrt{B}|W|
e^{\kappa K/2}K^{T \bar T}
{D_{\overline{T}}\overline{W}\over \overline{W}},\cr
  \partial_z ({1\over{\sqrt{ B(z)}}})&=
\kappa\zeta |W|e^{\kappa K/ 2}.} \eqn\bofz$$
with  the boundary condition that
at $z=Z$, \ie ,
at the location of the
domain wall  $B(Z)=1$. Eqs. \bofz\ and \summary\
 thus imply that
the line elements \metric\ and \pllineel\ have the
same conformal factors which
 depend on $r$ and $z$, respectively.
Clearly, for the choice of the
small solid angle in the direction of $\theta_0=0$,
one can identify  $z=r$ and the line elements
\metric\ and \pllineel\
are identical. This proves explicitly that the metric
solution obtained for the spherical domain wall in
the limit $R\to\infty$ is locally identical to the
metric of the planar domain wall.

We now turn to the explicit form for $B(r)$ in
specific cases. We study first the case with the
`false' vacuum being
 Minkowski ($W_{R+\Delta R}=0$) and
the bubble of `true' vacuum corresponding to the
AdS space ($W_{R-\Delta R}\neq0$).
In this case one obtains:
$$\eqalign{
B(r)\to1,&\hskip0.5cm      r>R+\Delta R;\cr
B(r)\to {3\over{\kappa|V_{true}|(r-R)^2}},&
\hskip0.5cm r<R-\Delta R,}
\eqn\madsm$$
where $V_{true}=-3\kappa|W|^2e^{\kappa K}\
_{|R-\Delta R}$.

 The second case corresponds to the solution with
`false' and `true'  vacua  both corresponding to
the AdS  supersymmetric vacua. In this  case
$W_{R-\Delta R}\neq 0$ and
 $W_{R+\Delta R}\neq 0$.
The asymptotic form of the metric is of the form:
$$ \eqalign{
B(r) \to  {1\over{[1-\sqrt{{\kappa|V_{false}|
\over 3}}(r-R)]^2}}, &\hskip0.5cm
 r>R+\Delta R, \cr
 B(r)\to{3\over{\kappa|V_{true}|(r-R)^2}},
&\hskip0.5cm r<R-\Delta R.}
\eqn\madsm$$
where $V_{(true, false)}=-3\kappa(|W|^2e^{\kappa
K})_{|R\mp\Delta R}$. The metric blows-up in the
region far outside the bubble
at $r = R + \sqrt {3 \over \kappa |V_{false}|} $,
but it is at an infinite proper distance away.

In this chapter,
we have calculated the Bogomol'nyi bound explicitly
for the energy density of the wall of a bubble
in the false vacuum  decay of a supersymmetric vacuum.
The calculation of the bound,
eqs. \minkowskibound\ and \adsbound,
is exact for  any value of $\kappa$.
The solutions to the first order Bogomol'nyi
equations, \ie\ those which saturate the
Bogomol'nyi bounds, were interpreted as describing a
bubble wall separating two distinct vacua.
The only solutions saturating the bound
correspond to supersymmetric configurations of the
bubble with its radius of curvature $R \to \infty$.
The analysis of the supersymmetric false vacuum decay
carried out by Weinberg\refmark{\WEIN} was to leading
order in $\kappa=8\pi G_N$ expansion (more
precisely $G_N |T|^2 << 1$). On the other hand,
by establishing the positive energy-density theorem
for the vacuum bubble we proved that the supersymmetric
vacua are nonperturbatively stable, \ie ,
there exists no false vacuum tunnelling to \sl all \rm
orders in $\kappa$.

It is intriguing that the Bogomol'nyi inequality
conditions \minkowskibound\ and \adsbound ,
derived from the positive energy-density theorem applied
to the vacuum decay bubble instanton configuration, is
exactly opposite to the Coleman-DeLuccia
bound \cdlbound\ and \cdlboundII\  for
allowing false vacuum decay. In other words, the
instanton for the tunnelling process between the
supersymmetric vacua is never materialized in supergavity
theories, including
the superstring vacua to which we now turn our attention.

\chap{Tunnelling Between Superstring Vacua}

In the previous section we found that a supersymmetric
vacuum is stable against tunnelling to an another
supersymmetric  AdS vacuum.
We now apply the result to the issue of vacuum
degeneracy in superstring compactifications.

In field theories, the inclusion of gravity usually
provides insignificant straightforward modifications
to the
matter sector of the theory. This is the case for the
tunneling process as well. Note that $\Delta V$
 in the
Coleman-DeLuccia bounds \cdlbound\ and \cdlboundII are
 proportional to $\kappa$ and thus
$\Delta V\to 0$ as $\kappa \to 0$.

In supergravity theories where matter fields
are associated with the low energy   scale, the
splitting of the non-degenerate supersymmetric vacua
is also generically small; \ie , $\prop \kappa$.
However, in superstring
theories, other fields, \eg , dilaton and  moduli,
and gravity are on an equal footing,
so the effects of gravity can yield
distinctly new features. In particular, when the vacuum
expectation values (VEV's) of the moduli fields   are
of the order of $M_{pl}$, the  non-perturbatively
induced potential is significantly modified by gravity.
In this case $\Delta V ={\cal O}(V)$ and prior to the
non-perturbative relations (eqs. \minkowskibound\
and \adsbound\ ) not much could be said about the
stability of such superstring vacua.

Moduli of superstrings compactified on Calabi-Yau
manifolds with a non perturbatively induced potential
are described by the $N=1$ supergravity
Lagrangian \lagran\ . As an example, we choose
the K\"ahler potential $K(T,\bar{T})$ and
superpotential $W(T)$ to be modular
invariant with the appropriate weights.  The
modulus field $T$ correspondings to the internal
size of the compactified space and acquires a non-zero
superpotential only
non-perturbatively, \eg\ via gaugino  condensation.
The symmetry of the Lagrangian reflects the underlying
target space modular invariance.

In section 2.1 we gave an example of a duality invariant
Lagrangian with one of the simplest  choices of the
superpotential, $W(T)=j(T)\, \eta^{-6}(T)$. In this case
the theory possesses a supersymmetric ground state (\ie ,
a point in field space satisfying $D_{T}W = 0$) at
$T=1$ ($W\neq 0$, AdS  space-time)  and
$T=\rho = e^{{i \pi \over 6}}$ ($W=0$, Minkowski
space-time). Recall that the value  of the
real part of the field $T$ describes an overall size
of the internal (compactified) space. Therefore,
$T=1$ and $T=\rho$ correspond to compactifications
of different size and shape.
Based on the result of the  previous chapter,
we can argue that there is \sl no \rm nonperturbative
instability of such superstring vacua.  In particular,
noncompact four-dimensional Minkowski space-time is stable
against decay into an AdS superstring vacuum. This  result
implies global stability of a supersymmetric vacuum in
string theory even in the case when the non-perturbative
physics removes degeneracy
due to the potential of the matter fields.
We can also
extend the above result also to the possibility of
decompactification instability. Tunnelling
from the local minimum $T=1$ or $T=\rho$ to $T\to\infty$
is impossible, since
the above choice of superpotential gives rise to a scalar
potential which diverges exponentially as $T \to \infty$.

There is a special choice of the
\REF\FONT{A.~Font, L.~E.~Ib\'a\~nez, D.~L\"ust,
and F.~Quevedo,    Phys. Lett. {\bf 245B} (1990) 401;
          S.~Ferrara, N.~Magnoli, T.~R.~Taylor, and
    G.~Veneziano, Phys. Lett. {\bf 245B} (1990) 409;
    P.~Binetruy and M.~K.~Gaillard, Phys. Lett. {\bf
    253B}(1991) 119;
    H.~P.~Nilles and M.~Olechowski, Phys. Lett. {\bf
    248B} (1990) 268.}
superpotential, best motivated in the study of gaugino
condensation\refmark{\FONT} \REF\Kaplun{V.~Kaplunovsky,
Nucl. Phys. {\bf B307} (1988)145.}
\REF\Dixon{L.~Dixon, V.~Kaplunovsky, and J.~Louis,
Nucl. Phys. {\bf B355} (1991) 649; J.~Louis,
{\it PASCOS 1991 Proceedings}, P. Nath ed.,
World Scientific  1991.}
 in a class of orbifold compactifications
with threshold
corrections \refmark{\Kaplun ,\Dixon}
included.  In this case
$W(T) = \eta^{-6}(T)$ and the potential
 has only
one AdS minimum at which the space-time
supersymmetry is spontaneously broken. The
 potential again diverges at $T \to \infty$. Thus,
decompactification instability does not exist in this
case either.

On the other hand, there are choices of modular
invariant superpotentials with vanishing scalar potential
$V\to 0$ as $T\to\infty$; for example, $W(T) =
\eta^{-6}(T)P(j(T))$ with the rational
polynominal $ P(j(T))= \sum_{n=1}^N \alpha_n j^{n}(T)/
\sum_{m=1}^M \beta_m j^{m}(T)$ and $M>N$.
\foot{See related discussion on the aymptotic behaviour
 and consequences
for the decompactification instabilities for a general
class of the duality invariant potentials in
Ref.~\refmark\CFILQ.}
However, these superpotentials in general possess,
for finite values of $T$, AdS minima with either
preserved or spontaneously broken supersymmetry.
Even in such cases, tunnelling to $T=\infty$ is
impossible. This is because $T \rightarrow  \infty$
is yet another point where the condition $D_{T}W \to
const.\times  W\to 0$ is satisfied, \ie , it is
yet another supersymmetric point.
Thus, superstring vacua
with such modular invariant potentials are also
stable against decompactification  to a
ten-dimensional flat space-time .

The nature of
the decompactifications pointed out above
does not quite correspond to
the non-compact Minkowski ten-dimensions. Since the
modular invariant superpotentials automatically
incorporate stringy target space duality symmetry,
the above discussed `decompactification' limit is
actually infinitely far away from the noncompact
Minkowski ten-dimensions in
the space of all possible superstring vacua.

 In the context of Kaluza-Klein
theory, the decompactification instability
 was shown in Ref.~\Ref\WIT{ E. Witten,
Nucl. Phys. \bf B195 \rm (1982) 41.} .
However, the instabilities in this case originate from
an \sl explicit \rm breaking of the spacetime
supersymmetry by twisted fermion boundary conditions
on the compactified directions.
Indeed, their results show \sl no \rm
instability at all if the compact direction is made
compatible with an existence of covariantly constant
spinors. On the other hand, these results should not be
applied as they stand to
nonsupersymmetric superstring compactifications unless
target space duality is properly taken into account
(Without taking into account duality symmetry, the
decompactification instability was discussed previously
\REF\MAZ{P. Mazur, Nucl. Phys. \bf B294 \rm
(1987) 525.}\REF\BH{D. Brill and G.T. Horowitz, Phys.
Lett. \bf B262 \rm (1991) 437.}
in Refs.~\refmark{\MAZ ,\BH}.)
Note that the target space duality symmetry is
insensitive to whether supersymmetry is broken or not
(spontaneously or explicitly), since the symmetry is
associated only with bosonic coordinate zero modes, while
none of the NS or R-sector fermions are involved (Recall
that R-sector fermion zero modes are defined at the
tangent spacetime coordinate patch.).
Furthermore, the winding numbers $N_w^a, \,\, a=5,
\cdots, 9$ around
each compactified directions are conserved quantum
numbers~\Ref\BV {R. Brandenberger and C. Vafa,
Nucl. Phys. \bf B316 \rm (1988) 316.}.
Since a true decompactification corresponds to
disappearance of any of these quantum numbers, we find it
impossible to change the global space topology. Also, the
winding modes guarantee an equivalence
relation between   small and   large radii of compactified
directions. This suggests that decompactification
instability of compactified directions is not possible
either,  and that no semiclassical instability exists
even in \sl explicitly \rm broken supersymmetric
superstring backgrounds.

{}From the above arguments, we come to an important conclusion
that there exist neither topology preserving nor topology
changing \sl semiclassical \rm instabilities in
four-dimensional
superstrings with a modular invariant superpotential.
Our work also complements the perturbative, local stability
argument of Ref.\refmark{\Dixonbanks} that
two local supersymmetric vacua are infinitely
far away, at least, for extended supergravity theories.

\chap{Discussion}

In this paper, we have investigated the issue of false
vacuum decay in supergravity theory, in particular,
in connection with the possibility of
settling to a unique superstring vacuum.
We have found that the supersymmetric vacua are stable
against false vacuum decay into other supersymmetric vacua
nonperturbatively, \ie , to all orders in a $\kappa=8\pi
G_N$ expansion. For example, the AdS supersymmetric vacua
are not connected via quantum tunnelling to the
supersymmetric Minkowski vacua.

Superstring vacua are known to give rise generically to a
multitude of local minima associated with the
nonperturbatively induced potential for the moduli fields.
Target space  duality symmetry for a class of string
backgrounds renders the superpotential modular invariant.
Clearly, a supersymmetric AdS local minima, which satisfies
$D_{T}W = 0, \,\,\, W \ne 0$, is lower
in vacuum energy than a Minkowski supersymmetric
local minima: $D_{T}W = W = 0$, by
an amount proportional to $\kappa=8\pi G_N$;
\ie , $V_{AdS}-V_{Mink}=
-3\kappa (|W|^2e^{\kappa K}_{|AdS}-
|W|^2e^{\kappa K}_{|Mink})$.
Thus, it appears that the AdS vacuum is the global
minimum  and one may expect the
Minkowski vacuum to tunnel to the AdS vacuum through
nonperturbative effects.

In this paper, we showed that such a
tunnelling is impossible. An
intuitive way of understanding our assertion follows from
consideration of the energy stored in the domain wall
of the bubble. The total energy is a sum of the matter
and the gravitational
energies. We found that the gravitational energy exactly
compensates that of the matter, thus making all the
local minima degenerate to one another.
The false vacuum tunnelling bounce solution ceases to exist
(equivalently, an imaginary part of the transition
amplitude vanishes, thus there is an infinitely long decay
time). The size of the bubble is infinite, since only a
planar, static domain wall is compatible with $N=1$
supergravity. In the case of superstring vacua, this leads
us to conclude that in a class of superstring vacua with
target space modular invariance the supersymmetric vacua
are stable against false vacuum decay.

In addition, we have established
the Bogomol'nyi bound for the energy density
 of the bubble wall  that corresponds
to the study of the false vacuum decay of a
supersymmetric AdS false vacuum to a supersymmetric
AdS true vacuum. Also in  such a case, saturation of
the Bogomol'nyi bound is precisely the limit that
the bubble surface tension is balanced by the energy
gained by tunnelling. This  constraint is given by
eq. \adsbound\ . Again, we found that the tunnelling
is completely suppressed, since the minimal energy
density stored in the domain wall is bound from
below and can at most saturate the Coleman-DeLuccia
bound (see eq.\cdlboundII).

If the supersymmetry of superstring vacua
is explicitly broken as in the case with twisted
boundary conditions in the internal space, the
positive energy theorem is not guaranteed, as shown
in Refs.\refmark{\MAZ ,\BH}. However, their analysis
is not fully stringy, since target space duality
invariance is not properly taken into account.
This invariance relates a small radius to a large
radius as an equivalence class, thus the
`decompactification' limit should be understood with
a full-fledged stringy consideration. Indeed, unless
winding modes around the compactified dimensions
disappear dynamically, conservation of winding
number suggests that the topology changing
decompactification, present in the
zero winding   sector instability, is impossible.

S.J.R. would like to thank G.T. Horowitz, E.W. Kolb,
E. Mottola, T.R. Taylor and A. Vilenkin for
enlightning discussions. M.C. and S.G. are grateful
to C. Burgess for clarifying  the
connection between spherical and Cartesian frames.
This work was supported in part by Department of
Energy and SSC
Junior Faculty Fellowship (M.C.),  Department of
Energy, National Science Foundation under Grant No.
PHY89-04035 and
Texas National Research Laboratory Commission (SJR).

\endpage

\Appendix{A}

The purpose of this Appendix is to present details
of the formulation of the positive energy bound and the
corresponding solution of the self-dual equations,
corresponding to a general spherically symmetric domain
wall configuration.
The general space time metric Ansatz is  of the form:
$$
ds^2 = B(r) dt^2 - A(r) dr^2 - C(r) d \Omega_2^2.
\eqn\metricApp
$$
The choice  $A(r) = B(r)$  can be taken without a loss of
generality by a redefinition of the radial coordinate.
In the context of the study of a vacuum bubble domain
wall  created at $t=0$, the  original $O(4)$ Ansatz
\ofourm\ lead us to the form \metric\ which
can be obtained from \metricApp\ with
$A(r) = B(r)$ and $C(r) = r^{2}B(r)$.

In this Appendix, we show
that no static compact domain wall
configuration exists. Namely,
static minimal energy spherical domain walls
exist only in the strict limit $R\to\infty$,
where $R$ is the radius of the spherical domain wall.
The metric the minimal energy domain wall has the
solution
$A(r) = B(r)$ and $C(r) = r^{2}B(r)$, \ie ,
the metric solution is the metric \metric\
of a vacuum bubble of an  infinite radius.

In the limit $R \to \infty$,
spherical domain walls
 flatten locally into planar domain walls.
Derivations and results presented here
should thus be compared to that of the planar domain
wall case studied in Ref.\refmark{\CGR}.

In the study of a large spherical domain bubble,
its  local region appears flat. As $R\to\infty$
the local analysis becomes exact and the physics is
independent of the local patch on the sphere chosen.
Then, the solid angle integral is immediate.
Yet, before the limit of flat wall is reached, it is
important to remember that in the local analysis
 we have a spherical and not planar topology, and thus
 the adjacent points  on the spherical bubble should
be probed. Therefore, when  probing  the $\hat{\theta}$
and $\hat{\phi}$ directions, the supersymmetry
spinor parameter $\epsilon$ as defined
in eq.\epsII\ is appropriate. On the other hand, when
probing points in the radial direction
the form \eps\ , rewritten here as
$$
\epsilon(r)^{T}
= B(r)^{1/4}(\alpha^{*}e^{{i\Delta(r) \over 2}},
                  \alpha e^{{i\Delta(r) \over 2}},
                  \alpha^{*}e^{-{i\Delta(r) \over 2}},
                 -\alpha e^{-{i\Delta(r) \over 2}}),
\eqn\epsApp$$
is appropriate. Here, $\alpha$ is an arbitrary complex
constant resulting from the residual `$N=1/2$'
supersymmetry which we give the value $|\alpha| = 1/2$ as
our normalization convention of the spinors.
This approach enables us to obtain the consistent form
of the Bogomol'nyi bound and the explicit solution
of the Bogomol'nyi equations in the limit $R\to\infty$.

As has become familiar in the analysis of supersymmetric
bosonic configurations, we start with the
construction of the Bogomol'nyi bound using a generalized
Nester's form.  Parallel with the analysis of
 Ref. \refmark{\CGR} , the
volume integral of the divergence of Nester's form is
related to the supersymmetry variation of the fermionic
fields in the theory in the following manner:
$$
\delta_{\epsilon}Q[\epsilon'] =
2\int_{\Sigma}\nabla_{\nu}N^{\mu \nu} d\Sigma_{\mu}=
\int C \sqrt{A \over B}
[-\delta_{\epsilon'}\psi^\dagger_i g^{ij}
\delta_{\epsilon}\psi_j + K_{T \bar T}\delta_{\epsilon'}
\chi^\dagger \delta_{\epsilon}\chi]dr d\Omega_{2}.
\eqn\volumeintegralApp$$
Here, the hypersurface $\Sigma$ has a measure
$d\Sigma_{\mu} = (d\Sigma_{t}, 0,0,0)$ where
$d\Sigma_{t} = |g|^{{1 \over 2}} drd\Omega_{2}$.
 For the metric Ansatz \metricApp\ , we have
$g = B^{2}C^{2}$.
By choosing  the Majorana four-spinors $\epsilon$ and
$\epsilon '$ equal,
the right hand side of \volumeintegralApp\  is
positive semi-definite.
The importance of this bound is due to the fact  that
\volumeintegralApp\ corresponds to the
 system's energy as can be seen
by the canonical algebra satisfied by the
supersymmetry charge $Q[\epsilon]$.

To obtain the ADM energy and topological charge for
the configuration, we work with the surface integral
form of \volumeintegralApp\ .
In order to obtain an explicit expression for the energy
and topological charge for the minimal configuration,
we must obtain solutions of the self-dual equations.
With the more general metric Ansatz \metricApp\ the
details are only slightly more complicated than in
Chapter 2, but the logic is the same.
We derive the self-dual equations from the conditions
that the supersymmetry variations on the fermions vanish.
The form of the equation arising from $\delta_{\epsilon}
\chi=0$ is the same: eq. \modulino\ .
For the gravitino, $\delta_{\epsilon}\psi_{\mu} = 0$,
we must construct the supercovariant derivative
$\hat{\nabla}_{\mu}$ as given by \gravitinotrans\ .

Again, the general expression for the supercovariant
derivative  acting on the spinor parameter $\epsilon$ is
$$\hat\nabla_{\rho}\epsilon \equiv
\delta_{\epsilon}\psi_{\rho} = [2\nabla_{\rho} +
ie^{\kappa K/ 2}(WP_{R} + \bar{W}P_{L})\gamma_{\rho}
 - Im(K_{T}\partial_{\rho}T)\gamma^{5}]\epsilon\ .
\eqn\gravitinotransApp$$
We assume the matter, geometry, and spinor to be time
independent as discussed in chapter 2.
The matter and geometry are furthermore only functions
of the radial coordinate $r$.
The natural orthonormal frame to use in constructing the
gravitational covariant derivative is a spherical frame
with diagonal Vierbein
$$
e^{\hat{t}} = B^{{1 \over 2}}dt, \  \
e^{\hat{\theta}} = C^{{1 \over 2}}d\theta, \  \
e^{\hat{\phi}} = sin\theta C^{{1 \over 2}}d\phi, \  \
e^{\hat{r}} = B^{{1 \over 2}}dr.
\eqn\frame$$
We find the following order for the Dirac matrices
convenient:
$$
\gamma^{t} = B^{-{1\over 2}}\gamma^{0}, \  \
\gamma^{\theta} = C^{-{1\over 2}}\gamma^{1}, \  \
\gamma^{\phi} = (C sin^{2}\theta)^{-{1\over 2}}\gamma^{2},
 \  \ \gamma^{r} = B^{-{1\over 2}}\gamma^{3}
\eqn\gammamatrices$$
where the numbered Dirac matrices are the usual flat space
matrices taken in a Weyl basis.

With these conventions and assumptions it is
straightforward to construct the supercovariant derivative.
Then, the first order differential equations are:
$$\eqalign{
\gamma^o \,
[\gamma^3 \, { \partial_r{\sqrt B} \over \sqrt B}
+ i\kappa
 {\sqrt B} e^{\kappa K/2}
(WP_R+\bar WP_L)
 ]\epsilon (r, \theta , \phi ) = 0, \cr
\{2 \partial_r - i \gamma^3 \kappa
{\sqrt B} e^{\kappa K/2}
(WP_R+\bar WP_L)  - \gamma^5\kappa
Im (K_T \, \partial_rT )\} \epsilon (r, \theta ,
\phi ) = 0,  \cr
\{ 2 \partial_\theta - \gamma^1 [\gamma^3
{\partial_r {\sqrt C} \over \sqrt B} +
i \kappa {\sqrt C} e^{\kappa K/2} (WP_R+\bar WP_L)]
   \} \epsilon (r, \theta , \phi )
= 0,\cr
 \{ 2 \partial_\phi - \gamma^2 [ \gamma^3 \sin \theta
({ \partial_r {\sqrt C} \over
\sqrt B}  + i\kappa {\sqrt C} e^{\kappa K/2}
(WP_R+\bar WP_L))
+ \gamma ^1 \cos \theta ]\} \epsilon (r, \theta ,
 \phi ) = 0.}
\eqn\gravitinoApp$$

Upon explicitly using the choice for the Dirac matrices
\gammamatrices\ and Majorana spinor $\epsilon$, one can
reduce the self-dual equations to simple scalar
relations to be satisfied by the matter, geometry,
and spinor.
Again, vanishing of the supersymmetry transformation \tofr\ of the
fermionic partner of $T(r)$  yields:
$$
\partial_{r}T(r) = i e^{i \Delta (r)} \sqrt A
K^{T \bar T}
e^{\kappa K/2} D_{\bar T} \bar W. \eqn\tofrApp
$$
The $t$-component of eq.\gravitinoApp\  yields
$$
  \partial_r({1\over \sqrt B})
= -i\kappa
e^{i\Delta (r)} e^{\kappa K/2}{\overline W}.
\eqn\tcompApp
$$
The $r$-component of
eq. \gravitinoApp\  constrains the
radial dependence of the spinor to be
$$
(\epsilon_{1}, \epsilon_{2}) =
  B^{1/4}  e^{i \Delta (r)/2}
\left(\tilde\epsilon_1(\theta , \phi ),
\tilde\epsilon_2
(\theta ,\phi ) \right), \eqn\epsApp$$
with
$$\tilde\epsilon_1(\theta ,\phi)
= \left(\tilde\epsilon_2 (\theta ,\phi)\right)^{*}.
\eqn\rcompApp$$
In addition, the phase $\Delta (r)$ satisfies
$$\partial_{r}\Delta (r) =
- \kappa Im(K_T\partial_rT). \eqn\deltar$$

With the expression for the metric
\tcompApp\ ,
the equation for the
$\theta$-component of \gravitinoApp\ can be written as
$$
[2 \partial_\theta - \gamma^1 \gamma^3
\partial_{r}\sqrt{{C\over B}}]\epsilon = 0
\eqn\thetacompApp
$$
while the $\phi$-component implies
$$
[2 \partial_\phi+
\gamma^1 \gamma^2 (-2 \sin \theta \partial_\theta
+ \cos \theta)] \epsilon = 0.
\eqn\phicompApp
$$
Now eq. \phicompApp\
can only be solved with a spinor of the form
$$
(\tilde\epsilon_1(\theta,\phi),\tilde\epsilon_2
(\theta ,\phi ))=e^{i\sigma_2\theta/2}e^{i\sigma_3
\phi/2}(a_1, a_2)
\eqn\epsIIApp$$
with $a_1$ and $a_2$ constant complex coefficients
and $\sigma_{2,3}$ the usual Pauli matrices.
With \epsIIApp\ for the spinor, eq.
\thetacomp\ implies the metric components satisfy
$$
\partial_{r}\sqrt{{C\over B}} = 1.
\eqn\CApp$$
With the boundary conditions   $B(R)=1$ and
$C(R)=R^2$, the solution of equation \CApp\ yields:
$$C(r)=B(r)r^2\eqn\solC $$
Thus, the  minimal energy solution for the
domain wall with the general metric Ansatz
\metricApp\ yields the metric configuration \metric\
of the $O(4)$ domain wall bubble at $t=0$.

On the other hand  the form for the spinors
\rcompApp\
and \epsIIApp\ are incompatible unless $R\to\infty$
limit is taken.
Thus, we are left with the same form of the geometry
we started
with in the $O(4)$ Ansatz in Chapter 2 as well as the
same conclusions in regard to the existence of static
spherical domain wall configurations;
{\it no static spherical compact configurations exist.}

We now turn to the calculation of the topological charge
and ADM mass density for the minimal configuration which
saturates the Bogomol'nyi bound.  For this purpose
we return to the Nester's form integral
\volumeintegralApp\ .
Using the anti-symmetry of Nester's form
yields
$$
 \int_{\Sigma}\nabla_{\nu}N^{\mu \nu} d\Sigma_{\mu}=
 \int_{\Sigma}\partial_{j}
(|g|^{{1\over 2}}N^{t j})drd\Omega_{2}
\eqn\stokesApp$$
where $j=\theta,\phi,r$.
Evaluating $N^{t j}$ consistent with the
approach   outlined shows that
only the term with $j=r$ contributes.
Separating the contributions due to gravity (the ADM
pieces associated with the gravitational covariant
derivative) from the purely matter pieces contributing
to the topological charge (terms proportional to
the $T(r)$ field), we have for the topological part
$$
 \int drd\Omega_{2} \partial_{r}(|g|^{{1\over 2}}
N^{t r}_{topo}) = 4\pi\int dr \partial_{r}[2C(r)
\zeta e^{\kappa K / 2} |W|].
\eqn\chargeIApp$$
The $4\pi$ arises from the $d\Omega_{2}$ integral.
In the thin wall approximation we can integrate over
the region $R-\Delta R$ to
$R+\Delta R$ and thus set $C(r)$ to its normalized value
$R^{2}$.   Dividing by the surface area $4\pi R^{2}$
allows us to identify  the topological charge per unit
area for the minimal configuration
$$
{\cal{C}} =
2(  \zeta
|We^{\kappa K \over 2}|)_{r=R+\Delta R}
-2( \zeta
|We^{\kappa K \over 2}|)_{r=R-\Delta R}\,
\eqn\chargeIIApp$$

This result is identical to that in the flat domain wall
case\refmark{\CGR} , as it    must be since \chargeIIApp\
is obtained in the flat limit: $R \to \infty$.
In this limit,
recall the thin wall approximation is exact.
Therefore, the energy
per unit area in the thin wall approximation for the
flattened bubble is $\sigma = |{\cal{C}}|$.

We now comment on the form of the topological charge
and the inequalities
\minkowskibound\ and \adsbound\ . First, it is important to
note that the explicit form of the topological charge
was obtained only for the supersymmetric configuration.
This charge is a real number.
These results should be contrasted with
the case in the corresponding globally supersymmetric
\REF\CQR{M. Cveti\v c, F. Quevedo and S.-J. Rey, Phys.
Rev. Lett. \bf 67 \rm (1991) 1836.}, \REF\AT{E. Abraham and P.
Townsend, Nucl. Phys. \bf B351 \rm (1991) 313.}
case\refmark{\CQR ,AT}.
There one knows the topological
charge without knowing the explicit minimal configuration
and this charge is complex.
One can trace the origin of the real
topological charge in the local case back to the
reality condition on the
space-time metric.
Namely, the reality of the metric imposes a phase relation
between the supersymmetry parameter $\epsilon$ and the
superpotential $W$ (see eqs.
\epsilont\ and \summary\ ).
The spinor parameter $\epsilon$ is therefore not a mere
neutral object in the extraction of the topological
charge as it is in the global $N=1$ theory.

Consider now a possible non-supersymmetric configuration
with non-trivial topology interpolating between two
supersymmetric vacua.
Its energy is strictly larger than its topological charge.
The question remains whether this charge can be less than
the known charge for the supersymmetric case.
We argue that the topological charge for this
configuration is greater or equal to that of the
supersymmetric configuration.
The non-supersymmetric configuration will not satisfy
the first order Bogomol'nyi (or self-dual)  equations
although it will satisfy the second order Euler-Lagrange
equations. The only form the topological charge can
have in this case is $\Delta(e^{i\gamma}|W|e^{\kappa
K /2})$, where $\gamma$ is some phase.
For the two cases of interest in this work; that of
tunnelling between
Minkowski to AdS, or AdS to AdS when $W$ never reaches a
zero, the topological charges have the form
$$
\eqalign{
C_{susy} &= |z_{1}| - |z_{2}|  \cr
C_{non-susy} &= z_{1} - z_{2} }
\eqn\minimalcharge$$
where $z_{i} = e^{i\gamma}|W|e^{\kappa K/2}_{|vacua}$.
It follows that
$$
\sigma_{susy} = |C_{susy}| \le |C_{non-susy}|
< \sigma_{non-susy}.
\eqn\chargebound$$
Therefore, the configuration which saturates the
Bogomol'nyi bound and interpolates between two
supersymmetric vacua is the minimal energy configuration.
Any statements concerning the energy $\sigma_{non-susy}$
of any other  bubble
wall interpolating between supersymmetric vacua
are thus bounded from below by the energy
$\sigma_{susy}$ of this minimal energy configuration.

\refout

\end